\documentclass[aps,prl,twocolumn,graphicx,showpacs]{revtex4-1}
\usepackage{amsmath,amssymb,amsfonts}
\usepackage{bbold}
\usepackage{color}
\usepackage{url}
\usepackage{graphicx}

\draft % marks overfull lines with a black rule on the right

\newcommand{\eq}[1]{Eq.~(\ref{#1})}
\newcommand{\noeq}[1]{~(\ref{#1})}
\newcommand{\Ref}[1]{Ref.~\cite{#1}}
\newcommand{\Refs}[1]{Refs.~\cite{#1}}
\newcommand{\noRef}[1]{~\cite{#1}}
\newcommand{\leb}[1]{\text{d}[#1]}
\newcommand{\dd}[1]{\text{d}#1}
\newcommand{\Det}{\text{det}}

\newcommand{\tr}{\text{tr}}
\newcommand{\id}{\mathbb{1}}
\newcommand{\IR}{\mathbb{R}}
\newcommand{\IC}{\mathbb{C}}
\newcommand{\IH}{\mathbb{H}}
\newcommand{\IK}{\mathbb{K}}

\newcommand{\diag}{\text{diag}}

\newcommand{\dagg}{^{\dagger}}

\newcommand{\egapmax}[1]{E^{(#1)}_p([0,t];p)}
\newcommand{\egapmin}[1]{E^{(#1)}_p([0,s];0)}

\begin{document}

\title{Limiting Statistics of the Largest and Smallest Eigenvalues in the Correlated Wishart Model }%Tracy-Widom Law for the Distribution of the largest and smallest eigenvalue in the Correlated Wishart Model}

\author{Tim Wirtz$^{1\dagger}$, Mario Kieburg$^{\ddagger 2}$ and Thomas Guhr$^{1}$}
\email{thomas.guhr@uni-due.de\\$^{\dagger}$tim.wirtz@uni-due.de\\$^{\ddagger}$mkieburg@physik.uni-bielefeld.de}
\affiliation{$^1$Fakult\"at f\"ur Physik, Universit\"at Duisburg--Essen, 47048 Duisburg, Germany\\$^2$Fakult\"at f\"ur Physik, Universit\"at Bielefeld, 33501 Bielefeld, Germany}

% \author{Tim Wirtz} 
% \affiliation{Fakult\"at f\"ur Physik, Universit\"at Duisburg--Essen, 47048 Duisburg, Germany}
% \email{tim.wirtz@uni-due.de}
% \author{Mario Kieburg}
% \affiliation{Fakult\"at f\"ur Physik, Universit\"at Bielefeld, 33501 Bielefeld, Germany}
% \email{mkieburg@physik.uni-bielefeld.de}
% \author{Thomas Guhr}
% \email{thomas.guhr@uni-due.de}
% \affiliation{Fakult\"at f\"ur Physik, Universit\"at Duisburg--Essen, 47048 Duisburg, Germany}

%\homepage[]{Your web page}
%\thanks{}
%\altaffiliation{}

\newcommand{\referee}[1]{{\color{red}#1}}
\newcommand{\we}[1]{{\color{blue}#1}}

\date{\today}

\begin{abstract}
The correlated Wishart model provides a standard tool for the analysis of correlations in a rich variety of systems. Although much is known for complex correlation matrices, the empirically much more important real case still poses substantial challenges. We put forward a new approach, which maps arbitrary statistical quantities, depending on invariants only, to invariant Hermitian matrix models. For completeness we also include the quaternion case and deal with all three cases in a unified way. As an important application, we study the statistics of the largest eigenvalue and its limiting distributions in the correlated Wishart model, because they help to estimate the behavior of large complex systems. We show that even for fully correlated Wishart ensembles, the Tracy-Widom distribution can be the limiting distribution of the largest as well as the smallest eigenvalue, provided that a certain scaling of the empirical 
eigenvalues holds.
\end{abstract}


\maketitle 

Time series analysis yields rich information about the dynamics but also about the
correlations in numerous systems in physics, climate research,
biology, medicine, wireless communication, finance and many other
fields~\cite{chatfield,Kanasewich,TulinoVerdu,Gnanadesikan,BarnettLewis,VinayakPandey,AbeSuzuki,Muelleretal,Seba,SanthanamPatra,LalouxCizeauBouchaudPotters,ple02}.
Suppose we have a set of $p$ time series $M_j$, $\ j=1,\ldots,p$ of $n$ ($n\geq p$) time steps each, which are normalized to zero mean and unit variance. The entries are real, complex, quaternion, \textit{i.e.} $M_j(t)\in \IR, \IC,\IH$, for $\beta=1,2,4$ and $t=1,\dots,n$. We arrange the time series as rows into a rectangular data matrix $M$ of size $p\times n$. The empirical correlation matrix of these
data,
\begin{align} 
C = \frac{1}{n} MM\dagg \ ,
\label{definitionCmatrix}
\end{align}
with $\dagger$ the Hermitian conjugation, is positive definite and either real symmetric, Hermitian, or Hermitian self-dual for
$\beta=1,2,4$ and measures the linear correlations between the time series.  

The largest and the smallest eigenvalue of a correlation matrix are highly relevant in many fields. In a simple, interacting dynamical system\noRef{GARDNERASHBY,May1972}, occurring in physics\noRef{MajumdarSchehr2014}, biology\noRef{pianka2011evolutionary}, chemistry\noRef{Feinberg1979}, ecology\noRef{StefanoSi2012}, \textit{etc.}, the cumulative distribution function of the largest eigenvalue estimates the probability to find the system in a stable regime\noRef{MajumdarSchehr2014}. In high dimensional statistical inference, linear principal component analysis is a method to reduce the dimension of the observations to ``significant directions''\noRef{Anderson2003}. Especially, the  largest eigenvalue corresponds to the most ``significant'' component \noRef{muirhead,Anderson2003,Johnstone,johnstone2001}. Another example is factor analysis, where the largest eigenvalue can be used to study common properties\noRef{Anderson2003}. The ratio of largest and smallest eigenvalue is important for the statistics of the 
condition number\noRef{Edelman1992,Edelmann1988}, in numerical analysis including large random matrices. In wireless communication, eigenvalue based detection\noRef{ZengLiang2009,ZengLiang2007,CardosoEtAl2008} is a promising  technique for spectrum sensing in cognitive radio. It utilize the statistics of the ratio of largest and smallest eigenvalue to estimate certain statistical tests \noRef{WeiTirkkonen2009,PennaEtAl2009I,PennaGarello2011}. The smallest eigenvalue is important for estimates of the \textit{error of  a received signal}\noRef{Burel02statisticalanalysis,ChenTseKahnValenzuela,UpamanyuMadhow} in wireless communication, for estimates in linear discriminant\noRef{LarryWasserman} as well as in principal component analysis\noRef{Gnanadesikan}, it is most sensitive to \textit{noise} in the data\noRef{Gnanadesikan} and crucial for the identification of \textit{single statistical outliers}\noRef{BarnettLewis}.  In finance, it is related to the \textit{optimal portfolio}\noRef{Markowitz}.

These examples show the considerable theoretical and practical relevance to study the distributions $\mathcal{P}^{(\beta)}_{\footnotesize \text{max}}(t)$,
$\mathcal{P}^{(\beta)}_{\footnotesize \text{min}}(s)$ of the largest, respectively, smallest eigenvalue. Both quantities can be traced back to gap probabilities, namely
\begin{align}
\mathcal{P}^{(\beta)}_{\footnotesize\text{max}}(t) &=  \frac{\dd{}}{\dd{t}} \egapmax{\beta}\  , \label{pmaxE} \\ \mathcal{P}^{(\beta)}_{\footnotesize\text{min}}(s) &= - \frac{\dd{}}{\dd{s}}  \egapmin{\beta}  \ , 
\label{pminE}
\end{align}
where $E^{(\beta)}_p([a,b];m)$ is the probability to find $m$ out of $p$ eigenvalues in the interval $[a,b]$.

This article has three major goals: First, we provide for the first time a framework to map a large class of invariant observables in correlated Wishart ensembles to invariant matrix models. Second, we explicitly apply this framework to the cumulative distribution function\noeq{pmaxE} of the largest eigenvalue and find an invariant matrix model. Third, we show that for a certain class of $C$'s, $p/n$ fixed and $n,p$ tending to infinity the largest, respectively, smallest eigenvalue are Tracy-Widom distributed.

The ensemble of random Wishart correlation matrices $WW\dagg/n$\noRef{muirhead,Anderson2003} consists of $p\times n$ model data matrices $W\in\text{Mat}_{p\times n}(\IK)$, where $\IK=\IR,\IC$ or $\IH$ for $\beta=1,2,4$, such that upon average $\langle WW\dagg/n\rangle= C$. Data analysis strongly corroborate, see \textit{e.g.} Refs.~\cite{TulinoVerdu,AbeSuzuki,Muelleretal,Seba,SanthanamPatra,LalouxCizeauBouchaudPotters,VinayakProsenBucaSeligman2014}, the Gaussian Wishart model\noRef{muirhead,Anderson2003}, 
\begin{align}
P(W|C) \sim \exp\left(-\frac{\beta}{2}\tr~WW\dagg C^{-1}\right) \ .
\label{WRMdistribution}
\end{align}
The matrix $WW\dagg/n$ is known as Wishart correlation matrix. The corresponding measure $\leb{W}$ and all other measures $\leb{\cdot}$ occurring later on are flat, \textit{i.e.}, the products of the independent differentials. Due to the invariance of $\leb{W}$, invariant observables depend on average solely on the distinct, always non--negative eigenvalues $\Lambda_j,$  $ j=1,\ldots,p$ of $C$ which are referred to as the empirical ones.  We arrange them in the 
diagonal matrix $\hat\Lambda=\Lambda\otimes\id_{\gamma_2}$ and introduce $\gamma_2=1$ if $\beta=1,2$ and $\gamma_2=2$ if $\beta=4$ and for later purpose $\gamma_1=2\gamma_2/\beta$, where $\id_N$ is the unity matrix in $N$ dimensions.

We consider an observable $\mathcal O(WW\dagg)$ which is invariant under an arbitrary change of basis and  $\mathcal O (WW\dagg)= \mathcal{O}(W\dagg W)$. This is a very weak assumption when studying the eigenvalue statistics of $WW\dagg$. We are interested in the average
\begin{align}
\label{eq:FourierApproachAverage}
\left<\mathcal O \left(WW\dagg\right)\right> &= K \int\leb{W}\mathcal O \left(WW\dagg\right) P(W|\hat \Lambda)~,
\end{align}
where the integration domain is $\text{Mat}_{p\times n}(\IK)$ and $K$ is a normalization constant. The non-triviality of the integral\noeq{eq:FourierApproachAverage} is due to a group integral of the form
\begin{align}
\begin{split}
  &\Phi_\beta(X,\hat\Lambda^{-1})= \\& \int\dd{\mu( V)}\exp\left(-\frac{\beta}{2}
   \tr V \left(X\otimes\id_{\gamma_2}\right) V\dagg \hat\Lambda^{-1}\right),
\end{split}
\label{gi}
\end{align}
where $\id_N$ is $N$ dimensional unit matrix, the integration domain is $\text{O}(p),\text{U}(p)$ or $\text{USp}(2p)$ for $\beta=1,2,4$, respectively, and $X=\diag(x_1,\dots,x_p)$ are the distinct eigenvalues of $WW\dagg=V\left(X\otimes \id_{\gamma_2}\right)V\dagg$. It is known as the orthogonal, unitary or unitary-symplectic Itzykson-Zuber integral\noRef{ItzyksonZuber}. For the unitary case only, it can be computed analytically and is given in a closed form \noRef{ItzyksonZuber,Balantekin2000,SimonMoustakasMarinelli}.

We replace the invariants of $WW\dagg$ in \eq{eq:FourierApproachAverage} by those of the $n\times n$ matrix $W\dagg W$. Thus, after introducing a $\delta$-function and replacing $W\dagg W$ by a  matrix in the same symmetry class, say $Q$, we find
\begin{align}
\label{eq:FourierApproachThreeFoldIntegralII}
\begin{split}
 \left<\mathcal O \left(WW\dagg\right)\right> = K \int\leb{H,Q} \mathcal{O} \left( Q\right) \exp\left(\imath \tr  H Q\right)\\\times\int\leb{W}\exp\left(-\imath\tr H W\dagg W\right)P(W|\hat\Lambda)~,
 \end{split}
\end{align}
where the integral of $Q$ and $H$ is over the set of real symmetric, Hermitian, Hermitian self-dual matrices of dimension $n\times n$ for $\beta=1,2,4$, respectively. The $H$ integral is the Fourier representation of the delta function. A detailed mathematical discussion will be given elsewhere\noRef{WirtzKieburgGuhr}. The advantage of this approach is that $H$ couples to $W\dagg W$ while $\Lambda^{-1}$ to $WW\dagg$, see \eq{WRMdistribution}. Hence the integral over $H$ is invariant under $H\to UHU^{-1}$ with $U$ orthogonal, unitary, and unitary symplectic for $\beta=1,2,4$ respectively. The remaining $W$ integral becomes a Gaussian integral over an $np$-dimensional vector with entries in $\IK$, yielding 
\begin{align}
\label{eq:FourierApproachSecondTrial}
 \left<\mathcal O \left(WW\dagg\right)\right> = \int\leb{H} \frac{\mathcal{F}_n\left[\mathcal{O}\right]\left(H\right)}{ \Det^{1/\gamma_1}\left(\id_{np\gamma_2} + \imath H \otimes\frac{2}{\beta}\Lambda\right)}~.
 \end{align}
In the expression\noeq{eq:FourierApproachSecondTrial}, we introduce the Fourier transform of the observable $\mathcal{O}$
 \begin{align}
 &\mathcal F_n\left[\mathcal O\right]\left(H\right) =\frac{1}{(2\pi)^{\mu}} \int\leb{Q}\mathcal O \left( Q\right) \exp\left(\imath\tr H Q\right)
 \label{eq:FourierApproachFourierTransformed}
\end{align}
where $\mu=n(n+1)/2,n^2,n(2n-1)$ is the number of real degrees of freedom of $Q$ for $\beta=1,2,4$, respectively. If we know $\mathcal F_n\left[\mathcal O](H\right)$, we can express the average\noeq{eq:FourierApproachAverage} as an invariant matrix integral. Thereby we completely outmaneuver the Itzykson-Zuber integral\noeq{gi}. 

We exploit this general observation to the statistics of the extreme eigenvalues. The gap probabilities in Eqs.~(\ref{pmaxE}) and\noeq{pminE} can be written as an ensemble averaged observable. We carry it out for \eq{pmaxE} only, since for \eq{pminE} it  works analogously. The joint eigenvalue distribution function derived from \eq{WRMdistribution} is
\begin{eqnarray}
P(X|\Lambda) &=K_{p\times n} \left|\Delta_p(X)\right|^{\beta} \Det^{\upsilon}X 
              \, \Phi_\beta(X,\hat\Lambda^{-1}) \ ,
\label{jpdfX}
\end{eqnarray}
with normalization constant $K_{p\times n}$, Vandermonde determinant $\Delta_p(X) = \prod_{i<j}(x_j-x_i)$ and $\upsilon = \beta(n-p +1 -2/\beta)/2$, see \Ref{WirtzGuhrI,WirtzGuhrII}. As $\Phi_2$ is known, in the complex case the joint probability distribution function provides a representation that can be handled analytically\noRef{SimonMoustakasMarinelli}. The highly
non--trivial part is the group integral \eq{gi}.
% \begin{align}
%   \Phi_\beta(X,\hat\Lambda^{-1}) &= \int\dd{\mu( V)}\exp\left(-\frac{\beta}{2} 
%    \tr V X V\dagg \hat\Lambda^{-1}\right) \ ,
% \label{gi}
% \end{align}
% where the integration domain is $\text{G}_p=\text{O}(p),\text{U}(p)$ or $\text{G}_p=\text{USp}(2p)$ for $\beta=1,2,4$, respectively. It is known as the orthogonal, unitary or unitary-symplectic Itzykson-Zuber integral. 
The gap probability to find all eigenvalues below $t$ can then be cast into the form 
\begin{align}
\begin{split}
\egapmax{\beta} &= K_{p\times n} \int\leb{X} \left|\Delta_p(X)\right|^{\beta}\Det^{\upsilon}X\\
                                &\times \prod_{i=1}^p\Theta\left(t\id_p-x_i\right) \, \Phi_\beta(X,\hat\Lambda^{-1}) \ ,
\end{split}
\label{definitionE}
\end{align}
where $\Theta(x_i)$ is the Heaviside $\Theta$-function of scalar argument. The Heaviside function of matrix argument is known in terms of an  Ingham-Siegel integral, see \Ref{Fyodorov2002} and references therein. It is unity if its argument is positive definite and vanishes otherwise.
% It possess a matrix representation, if  $A\in \text{Herm}_N^\beta$, then
% \begin{align}
%  \label{thetafunction}
% \Theta\left(A\right) \sim \int\limits_{\text{Herm}_N^\beta}\leb{H} \frac{\exp\left(\tr(\imath H+\id_{\gamma_2N})A\right)}{\det^{\alpha_\beta/\gamma_1}\left(\imath H + \id_{\gamma_2 N}\right)}~,
% \end{align}
% where $\alpha_\beta = n -1 + 2/\beta$, $\gamma_1 = 2$ if $\beta=1$ and $\gamma_1 = 1$ if $\beta=2,4$. 
Positive definiteness is an invariant property implying that the $\Theta$-function depends  on the eigenvalues $a_i$ of $A$ only, 
\begin{align}
\label{thetadecomposition}
\Theta\left(A\right)=\Theta\left(a\right) =\prod_{i=1}^N\Theta\left(a_i\right)~.
\end{align}
Since the integral\noeq{definitionE} is over the whole spectrum of $WW\dagg$, we express the gap probability as averaged $\Theta$-function,
\begin{align}
 \begin{split}
\egapmax{\beta} &= K_{p\times n} t^{np\beta/2} \int\leb{W}  P(W|\sqrt{t}\hat\Lambda) \\
                                &\times \Theta\left(\id_{\gamma_2p}-WW\dagg\right) \ ,
\end{split}
\label{FullWModelEmax}
\end{align}
where $W\in\text{Mat}_{p\times n}(\IK)$. Analogously, the gap probability\noeq{pminE} is given by
\begin{align}
\begin{split}
 \egapmin{\beta} = K_{p\times n} s^{np\beta/2}\int\leb{\widehat W}  P(\widehat W|\sqrt{s}\hat\Lambda) \\
 \times \Theta\left(\widehat W \widehat W \dagg-\id_{\gamma_2p}\right) \Det^{(n-p)/\gamma_1} \widehat W \widehat W\dagg\ ,
 \end{split}
\end{align}
with $\widehat W$ is a square $p\times p$ matrix .

To employ our approach to the gap probability\noeq{FullWModelEmax}, we choose the observable to be $\mathcal{O}(WW\dagg) = \Theta(\id_{\gamma_2p} - WW\dagg)$. The matrix $\id_{\gamma_2n} - W\dagg W$ has $p$ distinct eigenvalues that coincide with those of $\id_{\gamma_2p} - WW\dagg$ and $n-p$ distinct eigenvalues that are exactly one. Hence, using \eq{thetadecomposition} it is evident that $\mathcal{O}(WW\dagg)=\mathcal{O}(W\dagg W)$. To exchange the $Q$, $H$ and the $W$ integral, we shift the contour of $H$ by $-\imath\id_{\gamma_2n}$ and find the inverse Fourier-Laplace transform in \eq{eq:FourierApproachSecondTrial}
\begin{align}
\begin{split}
\mathcal  F_n\left[\Theta\right]\left(H-\imath\id_{\gamma_2n}\right)\sim\frac{\exp\left( \tr\left(\imath H + \id_{\gamma_2n}\right)\right)}{\Det^{\alpha/\gamma_1} \left(\imath H + \id_{\gamma_2n}\right)}~,
\end{split}
\end{align}
where $\alpha=n-1+2/\beta$. If we diagonalize $H=U\left(Y\otimes \id_{\gamma_2}\right)U\dagg$, where $U$ is in one of the three groups and $Y=\diag(y_1,\dots,y_n)$ is the matrix of distinct eigenvalues of $H$, we arrive at a remarkable, new expression for the gap probability\noeq{definitionE}
\begin{align}
\begin{split}
 &\egapmax{\beta} =K_{p\times n} \int\limits_{\IR^n}\frac{\leb{Y}\left|\Delta_n(Y)\right|^\beta}{\Det^{\alpha_\beta\beta/2} \left(\imath Y + \id_{n}\right)}\\&\times\frac{\exp\left( \gamma_2 \tr\left(\imath Y + \id_{n}\right)\right)}{\prod_{k=1}^p\Det^{\beta/2}\left(\id_{n} +\left(\imath Y + \id_{n} \right)2\Lambda_k/t\beta\right)}~.
\end{split}
\label{DiagGap}
\end{align}
Likewise, we derive an invariant matrix model for the gap probability\noeq{FullWModelEmax}
\begin{align}
\begin{split}
 &\egapmin{\beta} =K_{p\times n} \int\limits_{\IR^p}\leb{Y}\left|\Delta_p(Y)\right|^\beta\\&\times\frac{\mathcal{F}_p\left[ \Theta\left(Q -\id_{\gamma_2p}\right) \Det^{(n-p)/\gamma_1}Q\right](Y)}{\prod_{k=1}^p\Det^{\beta/2}\left(\id_{p} +\imath 2\Lambda_k Y /s\beta\right)}~.
\end{split}
\label{DiagGapEmin}
\end{align}
The Fourier integral can be done using the differential operator constructed in appendix B of \Ref{KieburgGroenqvistGuhr}, but the expression becomes cumbersome and we do not need these details for the following discussion. 

Both results\noeq{DiagGap} and\noeq{DiagGapEmin} have a $p$-fold product of determinants in the denominator in common. Due to the exponent $\beta/2$, this eigenvalue integral can be studied, at least for $\beta=2,4$, using standard techniques of random matrix theory. For $\beta=1$ standard methods do not apply as square roots of characteristic polynomials appear.
\begin{figure}[t!]
 \centering
 \includegraphics[width=0.48\textwidth]{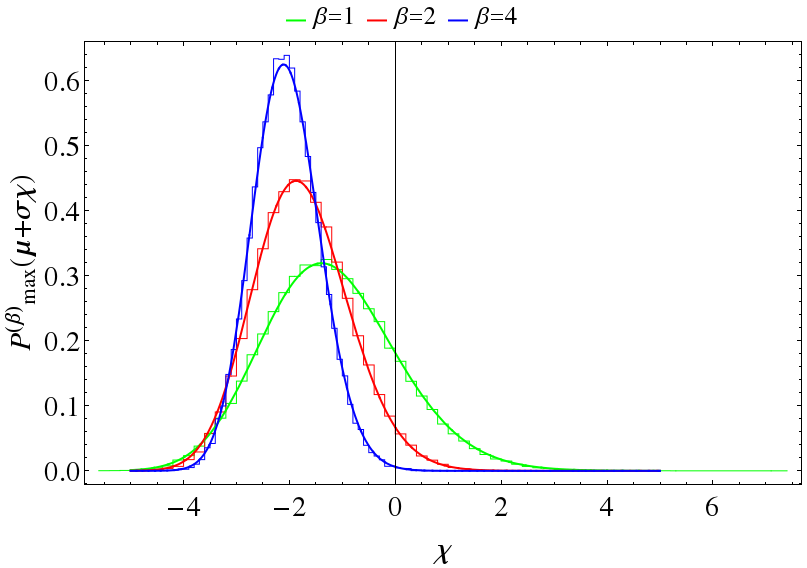}l%{./numerics-pmax-allbeta.png}
%  \includegraphics[width=0.48\textwidth]{./numerics-pmax-beta2.png}
%  \includegraphics[width=0.48\textwidth]{./numerics-pmax-beta2.png}
 % numerics-beta1.png: 1027x745 pixel, 96dpi, 27.17x19.71 cm, bb=0 0 770 559
 \caption{Comparison of analytic results for the largest eigenvalue distribution (straight lines) with numerical simulations (symbols) for $\beta=1,2,4$. We consider $80000$ realizations of $100\times300$ rectangular matrices, according to the distribution \eq{WRMdistribution}.}
\label{simul-largest}
\end{figure}
We further evaluate the exact expression\noeq{DiagGap} elsewhere\noRef{WirtzKieburgGuhr}. Here we focus on the limiting behavior which is more relevant in applications. To this end, we assume that the empirical eigenvalues are  random variables according to the distribution $\rho_{\text{emp}}(\Lambda)$ and $t =  \mu(\Lambda) + \sigma(\Lambda)\chi$, where the centering and scaling parameters $\mu(\Lambda)$ and $\sigma(\Lambda)$ are assumed to be large. We will show that this is a justified assumption.

For the uncorrelated Wishart ensemble, \textit{i.e.} $\Lambda=\id_p$, previous works\noRef{Edelman1992,Edelman1991,GuhrWettigWilke,Forrester1993709,DamgaardNishigaki,Johansson2000,johnstone2001,Soshnikov2002,Karoui2003,Karoui2006,VivoMajumdarBohigas2007,DeiftItsKrasovsky,Wang2009,FeldheimSodin2010,KatzavPerez,AkemannVivo2011} focus on the exact as well as the limiting distribution of the largest eigenvalue $x_{\text{max}}$ and the smallest eigenvalue  $x_{\text{min}}$ of $WW\dagg$. For $n,p$ tending to infinity, while $p/n=\gamma^2$ is fixed, it was proved that the limiting distribution of  $\chi_{\text{max}}=(x_{\text{max}}-\mu_{+})\sigma_{+}^{-1}$ and $\chi_{\text{min}}=(x_{\text{min}}-\mu_{-})\sigma_{-}^{-1}$ is the Tracy-Widom law $f_\beta(\chi)$ \noRef{tracywidomlong,TracyWidomPL,tracywidomGOEGSE}, where 
\begin{align}
\begin{split}
  \sigma_{\pm} = \pm\frac{(1\pm \gamma)^{4/3}}{\gamma^{1/3}}n^{1/3} ~~\text{and}~~
   \mu_{\pm} = (1\pm \gamma)^2n%(\sqrt{p} \pm \sqrt{n})^2~~,%\quad \tilde \mu_{n,p} =\left(\sqrt{p}-\sqrt{n}\right)^2 \\
%  \tilde \sigma_{p,n} &=  \left(\sqrt{p}-\sqrt{n}\right)\left(\frac{1}{\sqrt{p}}-\frac{1}{\sqrt{n}}\right)^{1/3}~, 
 \end{split}
 \label{eq:TW:JohnstoneScaling}
\end{align}
$\nu =n-p=(1-\gamma^2)n$ and $\gamma$ fixed for $p\rightarrow\infty$ \noRef{johnstone2001}. Moreover, if $n,p$ tend to infinity, while  $n-p$ is fixed it was shown that the limiting largest eigenvalue distribution is still Tracy-Widom\noRef{Forrester1993709,Johansson2000,johnstone2001}. 

For the correlated Wishart ensemble, the limiting largest eigenvalue distribution is known for $\beta=2$ in general\noRef{BaikBenArousPeche,ElKaroui2007} and for $\beta=1,4$ solely  when $\Lambda$ is a rank one perturbation of the identity matrix\noRef{Wang2009,MYMo,BloemendalVirag2013}. The smallest eigenvalue distribution was already studied in great detail in the microscopic limit, \textit{i.e.} $n,p\rightarrow\infty$ while $n-p=\nu$ fixed in \Refs{WirtzGuhrI,WirtzGuhrII}, whereas for $n,p\rightarrow\infty$ with $p/n$ fixed no results are available yet.

Similar to \Refs{WirtzGuhrI,WirtzGuhrII}, we assume that the empirical eigenvalues $\Lambda_k$ are of order $\mathcal O (1)$ for $n,p$ tending to infinity. It turns out that only the rescaled trace, $\left<\cdot\right>_{\text{s}} = p^{-1}\tr\left(\cdot\right)$, of $\Lambda^m$, where $m=1$, does not tend to zero. Moreover, another simple estimate shows $\left<\Lambda^m\right>_{\text{s}}  \sim \mathcal{ O} (1)$ such that we cannot determine the exact leading order of the empirical eigenvalue variance $\text{Var}_s(\Lambda)$. Consequently, we impose  another requirement on the empirical eigenvalue distribution, namely $ \text{Var}_{\text{s}}(\Lambda) \sim~\mathcal O (1/p^{\alpha})$, where $\alpha>0$ is a free parameter which we fix later. As a consequence of the Tschebyscheff inequality,
\begin{align}
 \mathbb{P}(\left|\Lambda - \left<\Lambda\right>_{\text{s}}\right| \geq x) \leq \frac{\text{Var}_{\text{s}}(\Lambda)}{x^2} \sim \mathcal O (1/p^{\alpha})~,\label{eq:TW:Tschebyscheff}
\end{align}
we make the following ansatz for the empirical eigenvalues by 
\begin{align}
\Lambda_{k} = \bar \Lambda + p^{-\alpha} \Lambda_{k}^{(1)}~,\label{lambdak}
\end{align}
where $\bar \Lambda =  \left<\Lambda\right>_{\text{s}} $ and $\Lambda_{k}^{(1)}\sim\mathcal{O}(1)$. If $C$ is a properly normalized correlation matrix, then $\bar \Lambda =  \left<\Lambda\right>_{\text{s}} =1 $.

Substituting \eq{lambdak} into  \eq{DiagGap}, and expanding the $p$-fold product to leading order, under the assumption that $t$ is large, we find for each  integration variable $y_i$
\begin{align}
\label{eq:TW:DeterminantExpansion}
\begin{split}
 &\prod_{k=1}^{p}\frac{1}{\left(1 + \left(\imath y_i + 1\right)\frac{2}{t\beta}\left(\bar \Lambda+p^{-\alpha} \Lambda_{k}^{(1)}\right)\right)^{\beta/2}}\\&=\left(1 + \frac{t\tr\Lambda^{(1)}}{p^{\alpha}p\bar \Lambda}\frac{\dd{}}{\dd{t}} +\dots\right) \frac{1}{\left(1 + \left(\imath y_i + 1\right)\frac{2\bar \Lambda}{t\beta}\right)^{p\beta/2}}~,
\end{split}
\end{align}
where $i=1,\dots,n$. The dots correspond to higher powers of $p^{-\alpha}$ times the derivative $t\dd{}/\dd{t}$. If we insert this expansion back into the cumulative distribution function\noeq{DiagGap} and keep only the leading terms in $n,p$ we find
\begin{align}
\begin{split}
 \egapmax{\beta}&=\left.\egapmax{\beta}\right|_{\Lambda=\bar \Lambda \id_p}\\&-\frac{\tr \Lambda^{(1)}}{p^{\alpha}p\bar \Lambda}t \frac{\dd{}}{\dd{t}}\left.\egapmax{\beta}\right|_{\Lambda=\bar \Lambda \id_p}+\dots~.
\end{split}
\label{eq:TW:ExpansionGapProbability}
\end{align}
The first term on the right hand side of \eq{eq:TW:ExpansionGapProbability} is \eq{definitionE} for  an uncorrelated Wishart ensemble with variance $\bar \Lambda$. From the discussion above \eq{eq:TW:JohnstoneScaling}, we conclude that if we center and rescale appropriately, the first term in \eq{eq:TW:ExpansionGapProbability} converges to the integrated distribution function $F_\beta(\chi)=\int_{-\infty}^\chi f_\beta(\chi')\dd{\chi'}$, found by Tracy and Widom. Therefore, we focus our discussion on the second term in \eq{eq:TW:ExpansionGapProbability}.

For the centered and rescaled threshold parameter $t=\mu_{+} \bar \Lambda +  \sigma_{+}\bar \Lambda \chi$, we take $t$ times the derivative with respect to $t$ of a function, which in the limit $n,p\rightarrow \infty$ and either $n/p$ or $n-p$ fixed converges to $F_\beta(\chi)$. Due to $p^{-1}\tr \Lambda^{(1)} \rightarrow \text{const}.$ for $p\rightarrow\infty$, the prefactor is of order $\mathcal{O}(p^{-\alpha})$. Solely the rescaling of the derivative $t\dd{}/\dd{t}$ can influence this order. A careful analysis shows if $\alpha$ is chosen such that for $n\rightarrow \infty$
\begin{align}
 \frac{1}{p^\alpha}\frac{\mu_{+}}{\sigma_{+}} = \gamma^{2\alpha-1}(1+\gamma)^{2/3} n^{2/3-\alpha}\rightarrow 0\ , %\sim\mathcal O \left(\frac{1}{n^{\alpha'}}\right),
\end{align}
the second term in \eq{eq:TW:ExpansionGapProbability} goes to zero as well. Thus, we require that $\alpha>2/3$ so that
\begin{figure}[t!]
 \centering
 \includegraphics[width=0.48\textwidth]{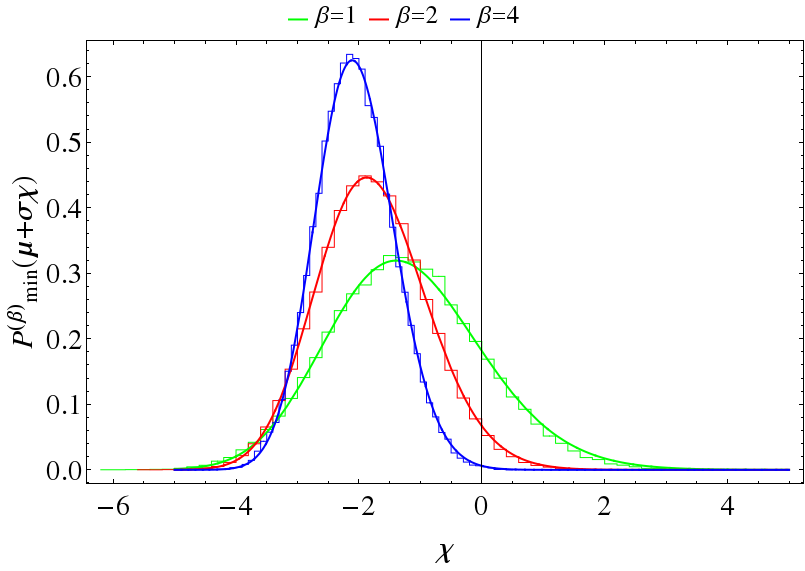}%{./numerics-pmin-allbeta.png}
%  \includegraphics[width=0.48\textwidth]{./numerics-pmin-beta2.png}
%  \includegraphics[width=0.48\textwidth]{./numerics-pmin-beta4.png}
 % numerics-beta1.png: 1027x745 pixel, 96dpi, 27.17x19.71 cm, bb=0 0 770 559
 \caption{Comparison of analytic results for the smallest eigenvalue distribution (straight lines) with the same numerical simulations (symbols) as in Fig.~\ref{simul-largest} for $\beta=1,2,4$. }
\label{simul-smallest}
\end{figure}
a macroscopic distance between the largest eigenvalue and the empirical eigenvalues is guaranteed only if the fluctuations of the empirical eigenvalues do not overlap with those of the largest eigenvalue.

Like the cumulative distribution function of the largest eigenvalue\noeq{DiagGap}, the dependence of the smallest eigenvalue gap probability\noeq{DiagGapEmin} on the empirical eigenvalues $\Lambda$ solely enter in the determinant in the denominator. Hence, we can apply the analysis done for the gap probability corresponding to the largest eigenvalue to that of the smallest one. Eventually, after centering and rescaling the threshold parameter, $s=\tilde \mu_{-}\bar \Lambda + \tilde\sigma_{-}\bar  \Lambda\chi$, where $\tilde \mu_{-}$ and $\tilde\sigma_{-}$ are as in \eq{eq:TW:JohnstoneScaling} and assuming the same restrictions on the empirical eigenvalue distribution as above, we obtain that cumulative distribution function is $F_\beta(\chi)$. 

To illustrate our findings, we compare our analytical results with Monte Carlo simulations for $\gamma\approx0.33$. The empirical eigenvalues are random variables with respect to a uniform distribution such that $\text{Var}_s\left(\Lambda\right) = p^{-7/4}$, $\left<\Lambda\right>_s=1=\bar\Lambda$ and $n^{2/3}\text{Var}_s\left(\Lambda\right) \approx 0.013\ll 1$. The  Comparison for the largest and the smallest eigenvalue distribution is shown in Fig.~\ref{simul-largest} and  Fig.~\ref{simul-smallest}, respectively. To demonstrate the agreement with the numerical simulations, we properly adjust the centering without changing the limit behavior. For the smallest eigenvalue we even properly adjust the scaling by a constant shift of the order $\mathcal O(1/n)$. This is because the smallest eigenvalue always ``feels'' the presents of a hard wall at zero, whereas the largest eigenvalue does not see any barrier such that the $1/n$ correction is stronger for the smallest eigenvalue.

In conclusion, we presented a new approach to map observables depending on the eigenvalues of a Wishart matrix only, to an invariant Hermitian matrix model. We demonstrated the concept by applying it to the gap probabilities corresponding to the largest and smallest eigenvalue distributions. Utilizing these invariant matrix model, we showed that for special empirical eigenvalue spectra, the Tracy-Widom distribution persist for the smallest and the largest eigenvalue if $n,p$ tend to infinity while $p/n=\gamma^2$ is fixed. We confirmed our findings by numerical simulations.

A simultaneous but independent study on related issues was very recently put forward in \Ref{KnowlesYin}.
\begin{acknowledgments}
We thank the Deutsche Forschungsgemeinschaft, Sonderforschungsbereich TR12 (T.W and T.G.) and the Alexander von Humboldt-Foundation (M.K.) for support.
\end{acknowledgments}

\bibliography{ref-maxeig-PRL.bib}{}

\end{document}